\documentclass[]{JHEP3} 

\usepackage{epsfig,multicol}

\newcommand\fverb{\setbox\pippobox=\hbox\bgroup\verb}
\newcommand\fverbdo{\egroup\medskip\noindent%
                        \fbox{\unhbox\pippobox}\ }
\newcommand\fverbit{\egroup\item[\fbox{\unhbox\pippobox}]}
\newbox\pippobox
\def\ie{{\it i.e.}}

\title{Next-to-leading order Higgs + 2 jet production via gluon fusion}

\author{John M. Campbell, \\
Department of Physics and Astronomy,\\
University of Glasgow, Glasgow G12 8QQ, UK\\
Email:~\email{j.campbell@physics.gla.ac.uk}}
\author{R. Keith Ellis, \\ 
Fermilab, Batavia, IL 60510, USA\\
and\\
PH Department, TH, CERN, 1211 Geneva 23, Switzerland\\
Email:~\email{ellis@fnal.gov}}
\author{Giulia Zanderighi\\
PH Department, TH, CERN, 1211 Geneva 23, Switzerland\\
Email:~\email{Giulia.Zanderighi@cern.ch}}

\preprint{\hepph{0608194}\\
CERN-PH-TH-06/164\\
FERMILAB-PUB-06-274-T}

\abstract{We present phenomenological results for the production 
of a Higgs boson in association with two jets at the LHC.
The calculation is
performed in the limit of large top mass and is accurate to
next-to-leading order in the strong coupling, \ie~${\cal O}(\alpha_s^6)$.}


\def\GeV{\mbox{GeV}}

\begin{document} 


\section{Introduction}

The search for the origin of electroweak symmetry breaking is the primary
goal of the LHC programme. This breaking can be realized by the introduction
of a Higgs boson whose expected physics signals and properties have been studied
extensively (for example, see Ref.~\cite{Djouadi:2005gi}). In particular
various search strategies have been devised in order to ensure that a Higgs
boson can be discovered over as large a range of putative masses as possible.

One of the main production mechanisms for the Higgs boson is the gluon fusion
process shown in Figure~\ref{fig:hprod}(a).
\FIGURE{\epsfig{file=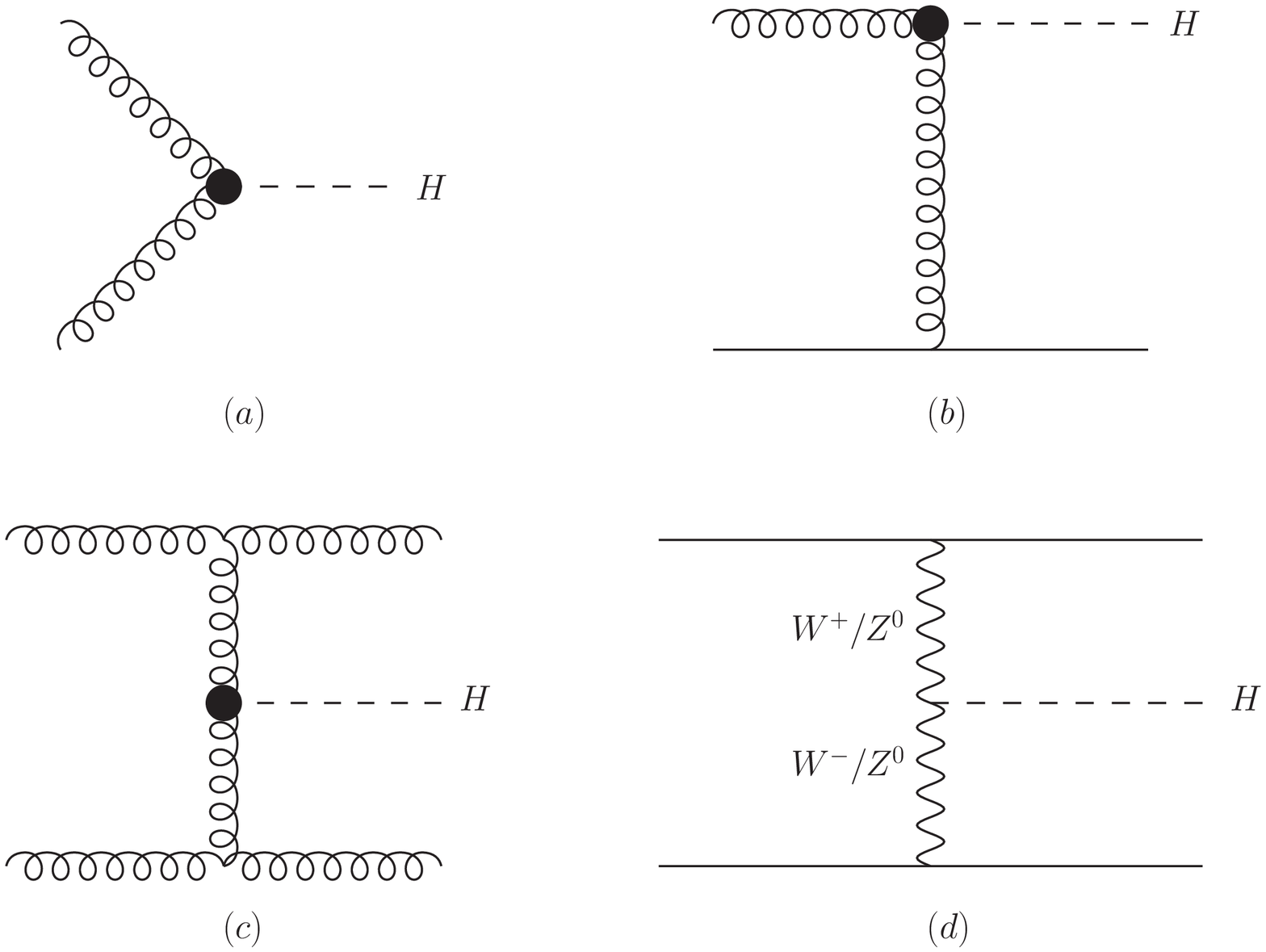,width=12cm,angle=0}
        \caption{Sample diagrams representing the production of a Higgs
        boson at the LHC. The basic gluon fusion process is represented in
        diagram (a), with an additional one and two hard partons in (b) and
        (c) respectively. Higgs production via weak boson fusion is
        depicted in diagram (d).}%
        \label{fig:hprod}}
In this diagram we have depicted
an effective coupling of the Higgs boson to gluons, which represents a top
quark loop in the limit of infinite top quark mass. The Lagrangian for the
effective theory is,
\begin{equation} 
{\cal L}_{\mbox{\rm eff}} = \frac{1}{4} A (1+\Delta) 
H G^a_{\mu \nu} G^{a\,\mu \nu}\, ,
\end{equation} 
where $G^a_{\mu \nu}$ is the field strength of the gluon field, $H$ is the
Higgs boson field and the effective coupling $A$ is given by 
\begin{equation} 
A = \frac{g^2}{12 \pi^2 v}\,.
\label{eq:Hggeff}
\end{equation} 
The effective coupling is thus dependent on the bare strong coupling $g$ and 
the vacuum expectation value parameter $v$, with
$v^2=(G_F\sqrt{2})^{-1}=(246~\GeV)^2$. 
The finite ${\cal O}(g^2$) correction to the effective operator
has also been calculated~\cite{Dawson:1990zj,Djouadi:1991tk}
\begin{equation} 
\label{Delta}
\Delta = \frac{11 g^2}{16 \pi^2}+ {\cal O}(g^4)\,.
\end{equation} 

The total cross section for the production of a Higgs boson via gluon
fusion at the LHC is of great importance.  Consequently not only have
the next-to-leading order (NLO) QCD corrections been
calculated~\cite{Dawson:1990zj}, but effects from one order beyond
that have also been
computed~\cite{Harlander:2002wh,Anastasiou:2002yz,Ravindran:2003um,Anastasiou:2005qj}.
In addition the process in which one additional hard parton is
observed (as in Figure~\ref{fig:hprod}(b)), so that the Higgs boson
acquires a non-zero transverse momentum at leading order (LO), has
also been calculated to
NLO~\cite{Schmidt:1997wr,deFlorian:1999zd,Ravindran:2002dc,Glosser:2002gm}.

It is then natural to consider the extension of this to the case in
which a Higgs boson is produced in association with two hard
partons~\cite{Kauffman:1997ix}.
There are in fact two types of processes that may contribute to such a
final state, the gluon-fusion process in Figure~\ref{fig:hprod}(c) and
the weak-boson fusion process in Figure~\ref{fig:hprod}(d). The latter
has been known to NLO for some
time~\cite{Han:1992hr,Figy:2003nv,Berger:2004pc}.  It is the
calculation of the full NLO QCD corrections to the Higgs$+2$~hard jets
process via gluon fusion in the large $m_t$ approximation that we will
present in this paper.

The lowest order amplitudes for Higgs$+2$~jet scattering have in fact
been calculated exactly, without using this effective
coupling~\cite{DelDuca:2001fn}. This is itself a $1$-loop computation
involving pentagon diagrams, so the calculation is complicated
considerably (and is currently intractable beyond this order).
However, it allows an examination of the limits in which the effective
coupling approach is valid. The results of this study indicate that
this approach is accurate as long as $m_H, p_T({\rm jet}) < m_t$, a
condition which will be satisfied for the results presented in this
paper.

\section{Structure of the calculation}

A detailed description of the calculation of the virtual matrix elements
using a semi-numerical approach is given in
Refs.~\cite{Ellis:2005zh,Ellis:2005qe}.

The real matrix element corrections to $H+2$~jet production are obtained by
including all crossings of results for the three basic processes,
\begin{eqnarray}
A) && 0 \to H q{\bar q} q^\prime {\bar q}^\prime g \; , \nonumber \\
B) && 0 \to H q{\bar q} ggg \; , \\
C) && 0 \to H ggggg \; . \nonumber
\end{eqnarray} 
For process $A)$ the relevant matrix elements of
Ref.~\cite{DelDuca:2004wt} are implemented. We have also used the
results therein for process $B)$ when the helicities of all gluons are
the same. For all other helicity assignments we have chosen to use the
more compact representation provided by the MHV techniques of
Ref.~\cite{Badger:2004ty}. Finally, process $C)$ exploits the NNMHV
matrix elements of Ref.~\cite{Dixon:2004za}, but uses the results of
Del Duca et al. for the other helicity combinations. Soft and
collinear singularities are handled using the dipole subtraction
scheme~\cite{Catani:1997vz}.

The calculation is incorporated into the general-purpose
next-to-leading order code MCFM.

\section{Results}

In order to render the cross section finite, we must apply some simple
cuts to the jets. The choice of cuts that we make is motivated partly
by the studies of Ref.~\cite{Berger:2004pc}, in which the sensitivity
of the Higgs cross section via the QCD and weak boson fusion processes
to the choice of minimum jet transverse momentum is studied.
For this choice of cuts, the cross section for the
production of a Higgs boson and two or more jets is dominated by the
$H+2$~jet contribution. As a result, the NLO QCD cross section for this
process shows the usual reduced dependence on the renormalization
and factorization scales.

\subsection{Inclusive cuts}

The simplest ``inclusive'' set of cuts that we consider is specified by
the following constraints on the jets, which are formed from the partons
according to the usual $k_T$-clustering algorithm~\cite{Blazey:2000qt}:
\begin{equation}
p_t({\rm jet})>40\;{\rm GeV}, \qquad
|\eta_{\rm jet}|<4.5, \qquad
R_{{\rm jet},{\rm jet}}>0.8 \; .
\label{eq:inclcuts}
\end{equation}
All of our results are based upon events in which at least two jets
satisfy these cuts, with the additional parton appearing at NLO
sometimes manifest as a third jet.  We do not consider the decay of
the Higgs boson and apply no cuts directly to the Higgs boson itself.

Before presenting any results, we note that the choice of 
parton distribution function (PDF) that is
used in the calculation is crucial. Each PDF set is
obtained by fitting a collection of observables with a particular value of
$\alpha_s(m_Z)$.  Since the effective coupling of the Higgs field to
two gluons is of order $\alpha_s$ (Eq.~\ref{eq:Hggeff}), the final
lowest order matrix elements squared for the $H+2$~jet process are
proportional to $\alpha_s^4$ and thus the cross section is very
sensitive to the input value from the PDF set. Throughout this paper
we will use two sets from the CTEQ6 package~\cite{Pumplin:2002vw}. At
leading order (LO) the CTEQ6L1 set is used (with $\alpha_s(m_Z)=0.130$
and $1$-loop running) whilst at NLO we have performed the calculations
with CTEQ6M ($\alpha_s(m_Z)=0.118$, $2$-loop running).

Higgs production via the weak boson fusion (WBF) process is most interesting for
``intermediate'' masses, in the region $115 < m_H < 160$~GeV. Since the QCD
production mechanism that we discuss in this paper is mostly of interest
as a source of additional events containing Higgs bosons in such a WBF
analysis, we limit our study to the same range. We choose the Higgs masses
at either end of the range and note that the effective $Hgg$ coupling that
we have used remains a good approximation for these values of $m_H$~\footnote{
It is possible to approximate the effects of performing the full calculation
at finite $m_t$ by scaling all our results by the ratio of the leading order
cross sections in each approach, according to the results of
Ref.~\cite{DelDuca:2001fn}. Since the difference is at most only a few percent
for $m_H=160$~GeV, we have not done so here.}.

\TABULAR{lll}{
Higgs mass            & $115$~GeV   & $160$~GeV   \\
\hline
 $\sigma_{\rm LO}$ [pb]   & $3.50$      & $2.19$ \\
 $\sigma_{\rm NLO}$ [pb]  & $4.03$      & $2.76$ \\
\hline
 $\sigma_{\rm WBF}$ [pb]  & $1.77$      & $1.32$
}
{LO and NLO cross sections for the gluon-fusion process with the basic
inclusive cuts of~(Eq.~\ref{eq:inclcuts}), together with the weak boson
fusion cross section at NLO.
\label{tab:inclxsec}}
Our results for the cross sections with the cuts given
by~Eq.~(\ref{eq:inclcuts}) are summarized in Table~\ref{tab:inclxsec}.
Cross sections for the weak boson fusion process
at NLO are also shown in Table~\ref{tab:inclxsec}.
They are about half the size of those for the gluon fusion process.
The default value for both the renormalization and
factorization scales is $m_H$. With this choice of scale and
the cuts of Eq.~\ref{eq:inclcuts}, the NLO corrections to the
$H+2$~jet cross section are quite mild, increasing the LO cross
section by only $15\%$ for the lower mass and by $26\%$ for a
$160$~GeV Higgs boson.
Although the corrections to the $H+2$~jet cross section via gluon fusion
are a little larger than those for the corresponding WBF process
(which are around $10\%$), existing LO analyses would not be much
changed by the inclusion of NLO corrections via such an inclusive
$K$-factor. In contrast, the $H+1$~jet cross section,
using $m_H=115$~GeV and the same choice of jet definition,
increases by about $50\%$ at NLO.

The dependence of the cross section on the unphysical renormalization
and factorization scales can be used to assess not only the
sensitivity of physical predictions to these inputs, but also the
trustworthiness of the perturbative series itself. To that end, we
show in Figure~\ref{fig:mudepincl} the scale dependence obtained from
our LO and NLO predictions, for each choice of Higgs mass.
\FIGURE{\epsfig{file=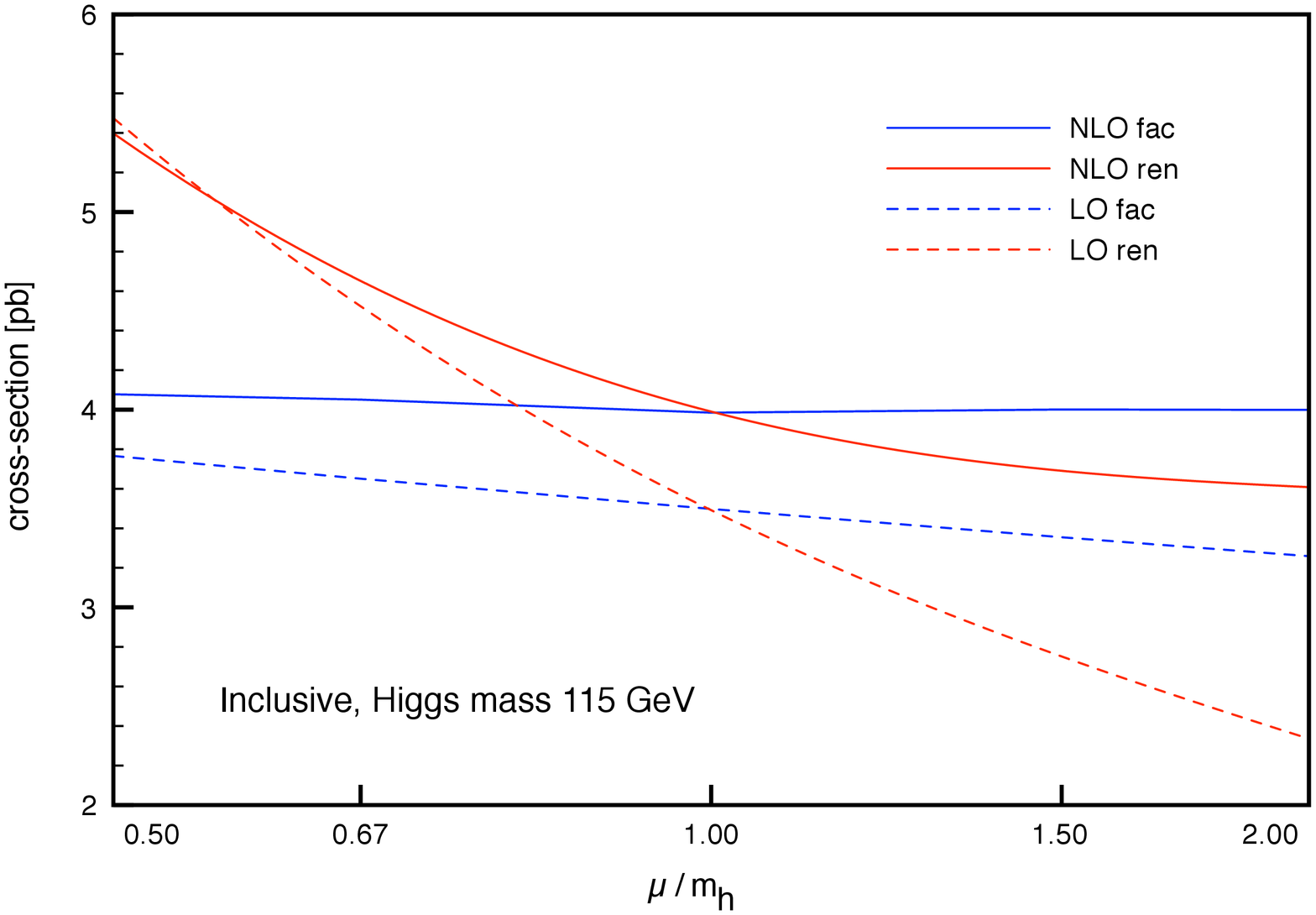,width=12cm,angle=0}
        \epsfig{file=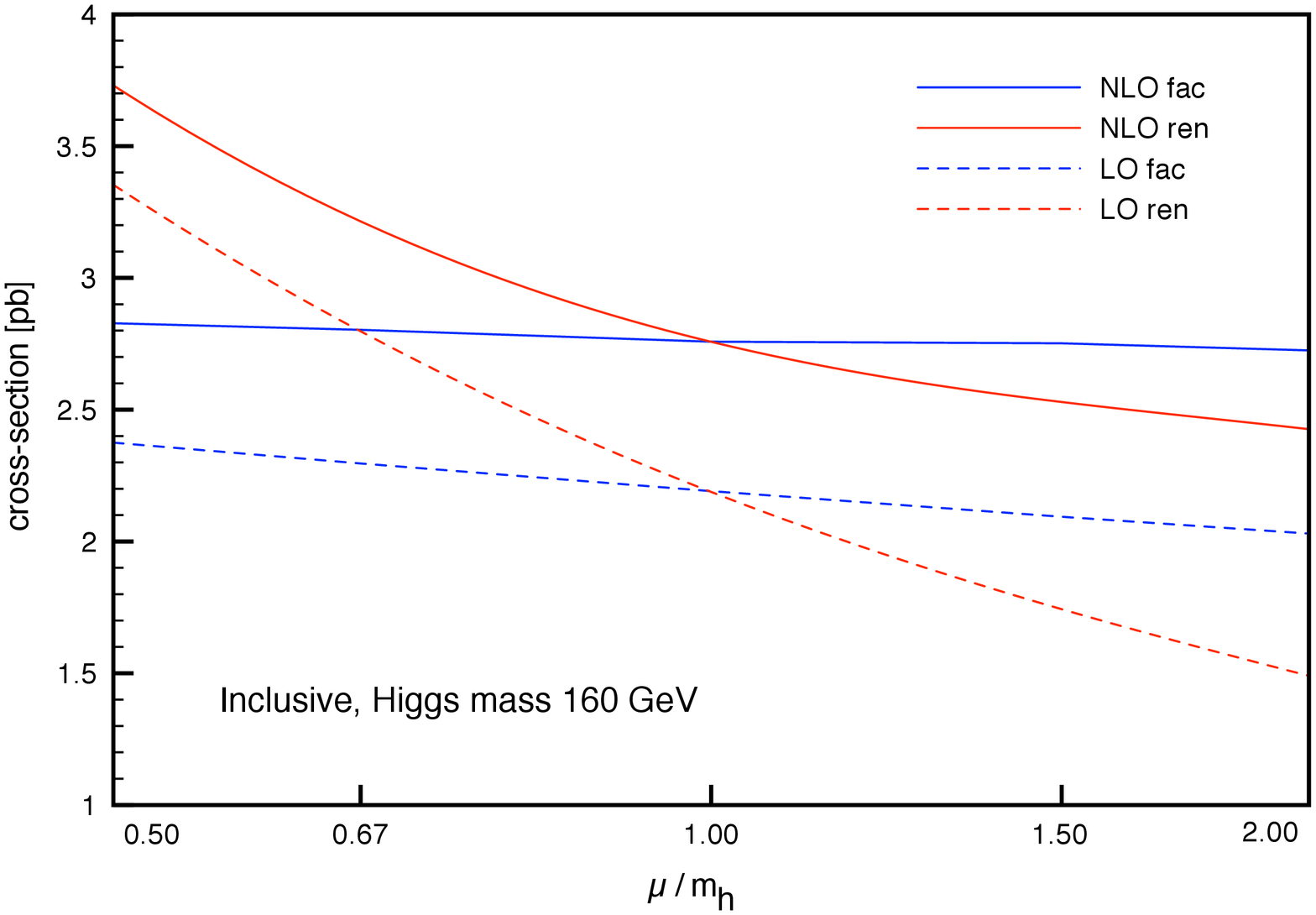,width=12cm,angle=0} 
        \caption{Scale dependence of the Higgs$+2$~jet cross section
        with minimal rapidity and transverse momentum cuts, for
        $m_H=115$~GeV (upper) and $m_H=160$~GeV (lower).}%
        \label{fig:mudepincl}}
The dependence on each scale is shown separately, with a
variation by a factor of two about the default value of $m_H$,
while the other scale is held fixed. The shapes of the curves
for the two Higgs masses are very similar. In each case the
dependence of both the LO and NLO cross sections on the
factorization scale is very mild, with a slightly smaller
variation at NLO. In contrast, the dependence on the
renormalization scale is very significant at LO, as expected.
This is somewhat reduced at NLO, although a large dependence
remains, with the cross section increasing by about $35\%$ when
the renormalization scale is halved. Therefore the NLO cross
section still contains a significant residual uncertainty that
should be accounted for in an honest analysis. This intrinsic
lack of reliability is in stark contrast to the WBF process,
which contains very little dependence on the input scales. This
is demonstrated in Figure~\ref{fig:wbfproc}, where we show the
behaviour of the NLO cross section for production of a Higgs
boson by WBF, under the same set of cuts.
\FIGURE{\epsfig{file=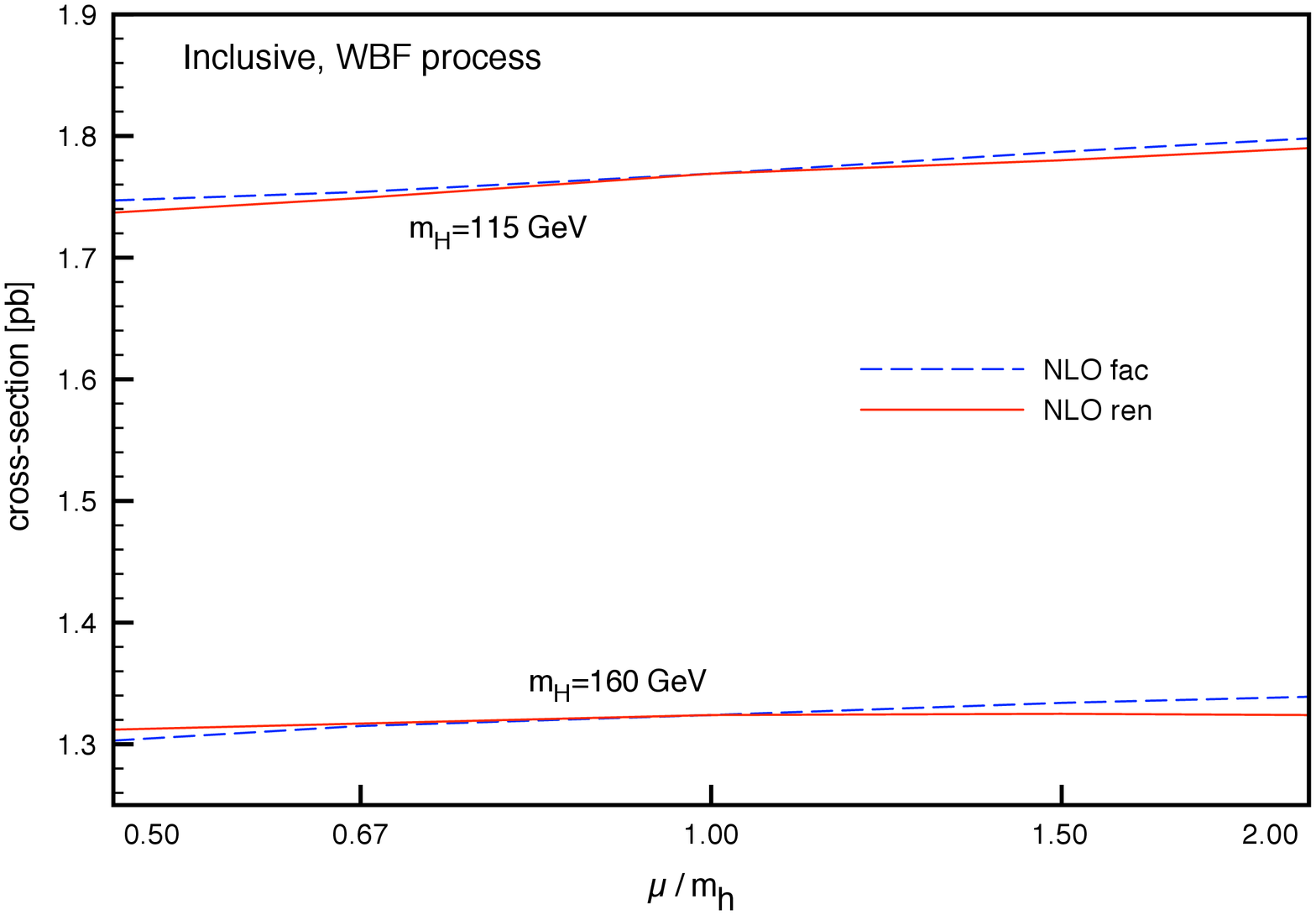,width=12cm,angle=0}
        \caption{Scale dependence of the weak boson fusion Higgs cross section
        with minimal rapidity and transverse momentum cuts, for $m_H=115$~GeV
        (upper set of curves) and $m_H=160$~GeV (lower).}%
        \label{fig:wbfproc}}
For both masses, the cross section varies only by approximately
$1.5\%$ over the same range of scales as considered above.
      
We note that a choice for the minimum jet transverse momentum
lower than given in Eq.~(\ref{eq:inclcuts}) results in similar
scale-dependence plots, but with much steeper renormalization
scale curves. We interpret this as a sign that the perturbative
series is less well-behaved for such a choice of cuts. Indeed,
in those cases, the cross section for producing a Higgs$+3$~jet
final state becomes larger than the one for Higgs$+2$~jets.

The calculation that we have performed can be used to study much
more than the two-jet inclusive cross section above. In the
analysis of Higgs production via weak boson fusion, it is
imperative to study the distribution of the rapidities of the
jets that are produced and in particular, whether or not this is
changed by NLO corrections. It is clearly prudent to perform the
same analysis for the process at hand.
In Figure~\ref{fig:etaincl} we present the results of such a study.
\FIGURE{\epsfig{file=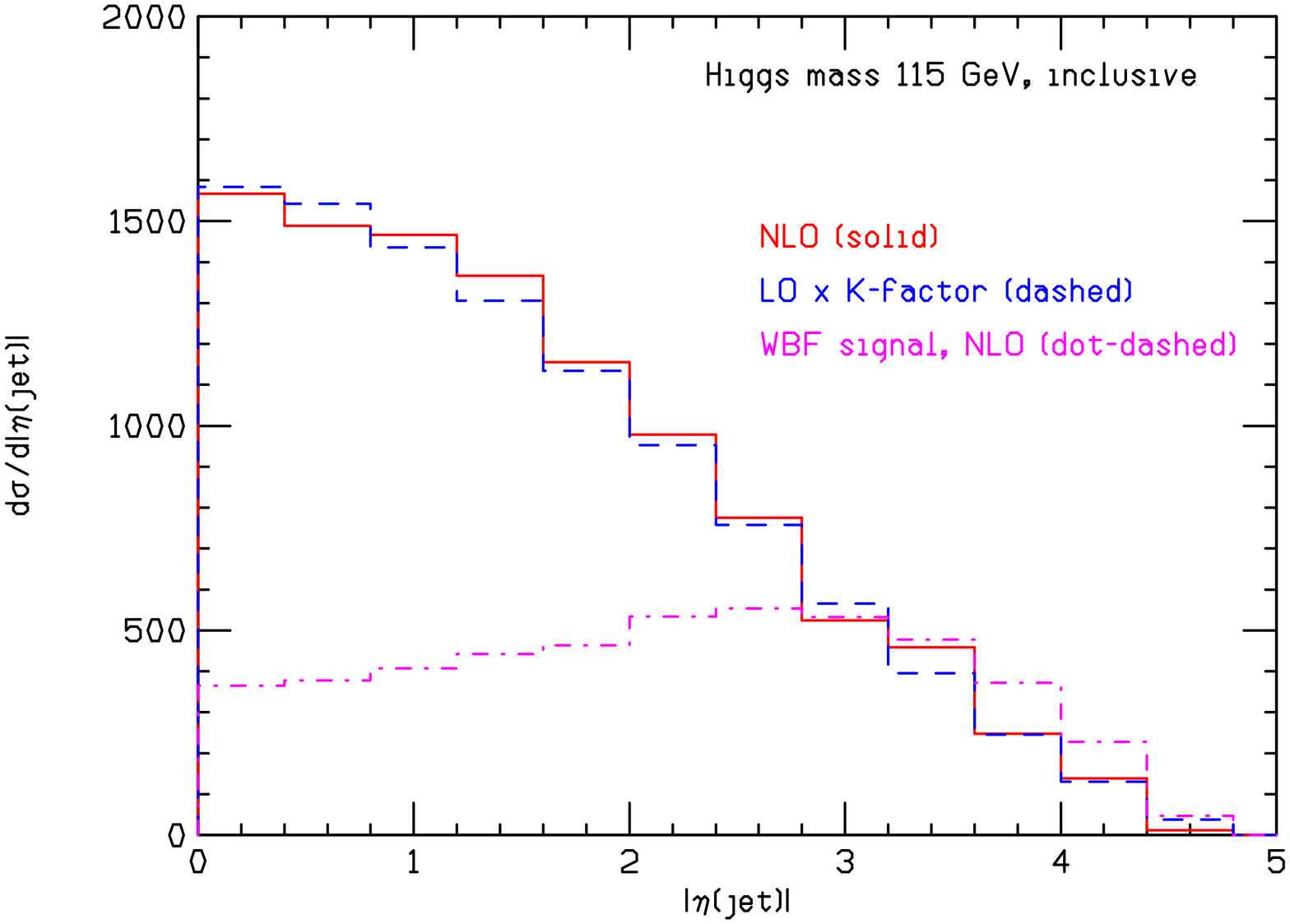,width=8cm,angle=90}
        \epsfig{file=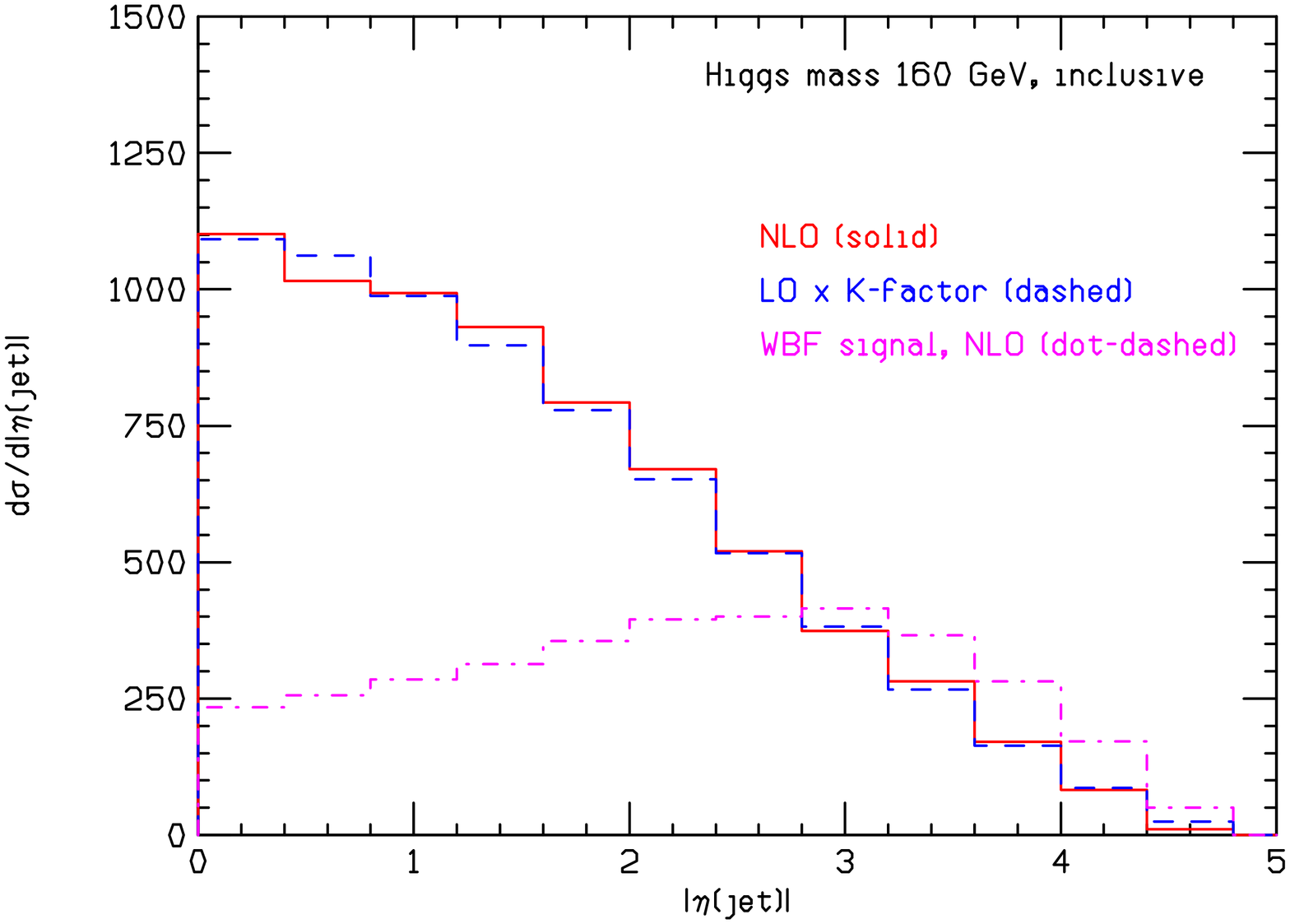,width=8cm,angle=90} 
        \caption{The pseudorapidity distribution of the two leading jets
          using only the inclusive cuts, for $m_H=115$~GeV (upper) and
          $m_H=160$~GeV (lower). The QCD process is calculated at LO
          (rescaled by the inclusive $K$-factor) and at NLO. The NLO
          result for weak boson fusion is also shown for comparison.
          Both jets in an event enter the
        histogram, each with weight one half.}%
        \label{fig:etaincl}}
In this study the two jets in each event with the largest transverse
momenta are chosen. Their rapidities are entered into the histogram with
weight one half, so that the area under the histograms yield the NLO cross
sections shown in Table~\ref{tab:inclxsec}. The LO distribution
has been scaled up by the ratio of cross sections in that table, so that its
area is the same. The distribution shows no evidence for any change
of shape when including the QCD corrections.
The figure also shows the NLO prediction for the weak boson fusion process,
which is significantly different. Even with this very minimal set of
cuts, it is clear that requiring just one of the two jets to be produced at a
relatively large rapidity (for example, $2$ units) significantly enhances the
production of Higgs bosons via WBF with respect to the QCD production
mechanism.

\subsection{Weak boson fusion cuts}

In addition to the basic cuts discussed above, we have also considered
a set of cuts that is designed to enhance the weak boson fusion Higgs
process and suppress processes which involve the production of
additional jets via QCD.  There are many variants of these cuts, all
designed to pick out configurations involving one or more forward
jets. We choose a fairly minimal set of constraints on the two jets
with the highest transverse momenta ($j_1$ and $j_2$), in addition to
the cuts in Eq.~\ref{eq:inclcuts}, we impose
\begin{equation}
|\eta_{j_1}-\eta_{j_2}|>4.2, \qquad \eta_{j_1}\cdot \eta_{j_2}<0 \; .
\label{eq:wbfcuts}
\end{equation}
Thus the two ``tagging jets'' are required to be both well-separated in
rapidity and to lie in opposite hemispheres. Note that we have refrained from
using the term ``rapidity gap'' to describe this separation, since the
additional softer parton that can be present at NLO may lie between the
two tagging jets.

\TABULAR{lll}{
Higgs mass            & $115$~GeV   & $160$~GeV   \\
\hline
 $\sigma_{\rm LO}$ [fb]   & $271$       & $172$ \\
 $\sigma_{\rm NLO}$ [fb]  & $346 \pm 5$ & $236 \pm 3$ \\
\hline
 $\sigma_{\rm WBF}$ [fb]  & $911$       & $731$
}
{LO and NLO cross sections with the weak boson fusion search cuts
of Eq.~(\ref{eq:wbfcuts}). Errors on the NLO cross sections are statistical
only. Also shown are the WBF cross sections at NLO.
\label{tab:wbfxsec}}
The cross sections that we find when using this set of cuts are shown
in Table~\ref{tab:wbfxsec}. We again choose both renormalization and
factorization scales equal to $m_H$ for these predictions. From the
lowest order cross sections it is clear that this set of cuts probes a
rather small fraction, less than $8\%$, of the total cross section in
Table~\ref{tab:inclxsec}. For this reason it is somewhat harder to
perform the integration over the phase space at NLO and we have thus 
indicated the statistical errors from our numerical integration in the
table.  These errors are at the level of $1.5\%$.  With these cuts,
the effect of the QCD corrections is greater than in the more
inclusive case. The cross section increases by about $30\%$ for
$m_H=115$~GeV and by a little more, around $40\%$, for the higher
mass. In this table we also show the corresponding cross sections for
the weak boson fusion process, for comparison. One can see that the
cuts of Eq.~(\ref{eq:wbfcuts}) have been quite effective in
suppressing the gluon-fusion process, so that the cross sections due
to weak boson fusion dominate by about a factor of $2.5$.

We also repeat the scale dependence study that was performed with the
inclusive cuts in the analogous Figure~\ref{fig:mudepwbf}.
The results are broadly similar, with the cross section showing little
factorization scale dependence but a considerable variation with the
renormalization scale. Again this renormalization scale dependence is
decreased somewhat at NLO, but a large uncertainty on the cross
section remains.
\FIGURE{\epsfig{file=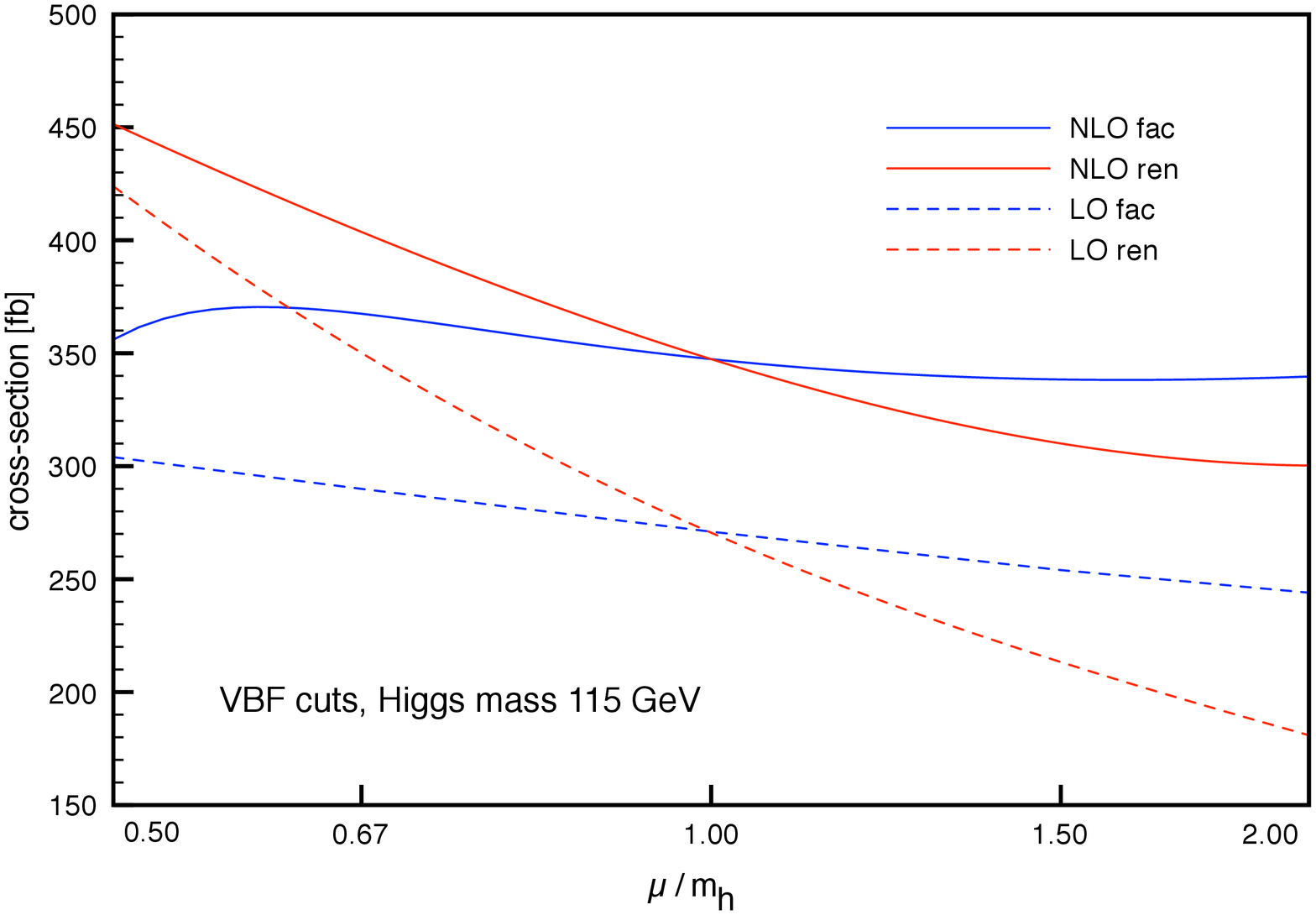,width=12cm,angle=0}
        \epsfig{file=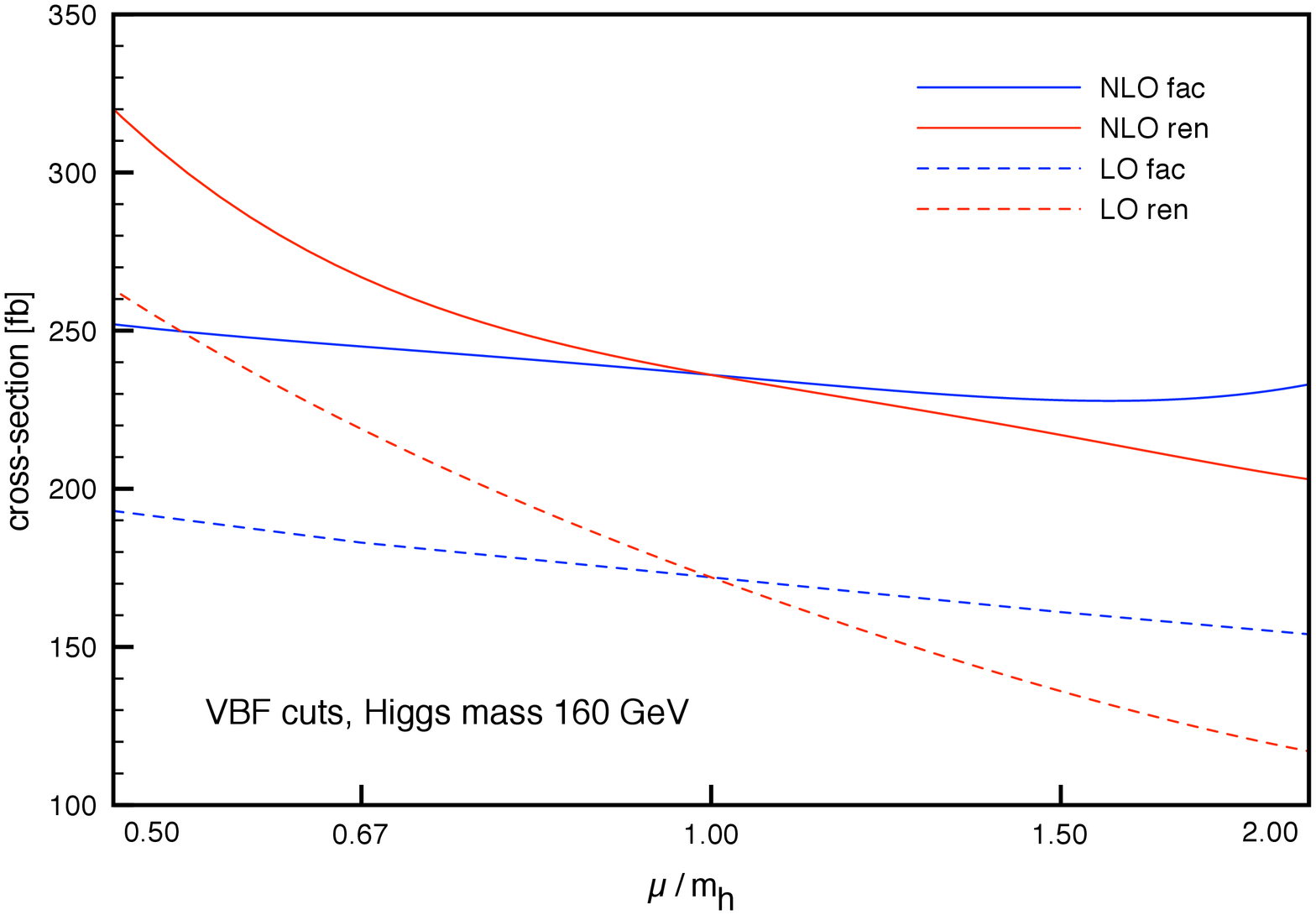,width=12cm,angle=0}  
        \caption{Scale dependence of the Higgs$+2$~jet cross section
        in the region selected by the weak boson fusion cuts, for
        $m_H=115$~GeV (upper) and $m_H=160$~GeV (lower).}%
        \label{fig:mudepwbf}}

Of particular interest to the Higgs boson search is the distribution of the
azimuthal angle between the two accompanying jets ($\Delta\phi$).
Because of the CP-even (scalar) nature
of the Higgs boson, the QCD process which we consider in this paper produces
a large correlation between these two jets, with a $\Delta\phi$ distribution
that is peaked at $0$ and $\pi$ and heavily suppressed at $\Delta\phi = \pi/2$.
This is in contrast to the weak boson fusion process, which produces an almost
flat distribution in $\Delta\phi$. Therefore this observable has been
considered as an additional discriminator between the two processes. Moreover,
a CP-odd (pseudoscalar) Higgs boson would produce a very different
distribution in the QCD
process (with the position of the peaks and trough reversed), so that this
mechanism could potentially be used to probe the CP properties of the Higgs.

These observations pertain to the lowest order predictions. It is
natural to consider whether or not they still apply in more detailed
studies.  An investigation of the effects of a parton
shower~\cite{Odagiri:2002nd} suggested that $\Delta\phi$ is subject to
logarithmically-enhanced higher order corrections and the correlation
is reduced. A more recent study, aiming to separate the effects of
hard radiation from those of showering and
hadronisation~\cite{DelDuca:2006hk}, finds that the correlation
largely survives both effects. With the calculation presented in this
paper we are able to provide further insight by studying the effects
of the NLO corrections to this observable.

The results of our study are shown in Figure~\ref{fig:phi56wbf}, where
we have again scaled up the lowest order prediction by the ratio of
cross sections in Table~\ref{tab:wbfxsec}, so as to enable a
comparison of the shapes of this distribution.
\FIGURE{\epsfig{file=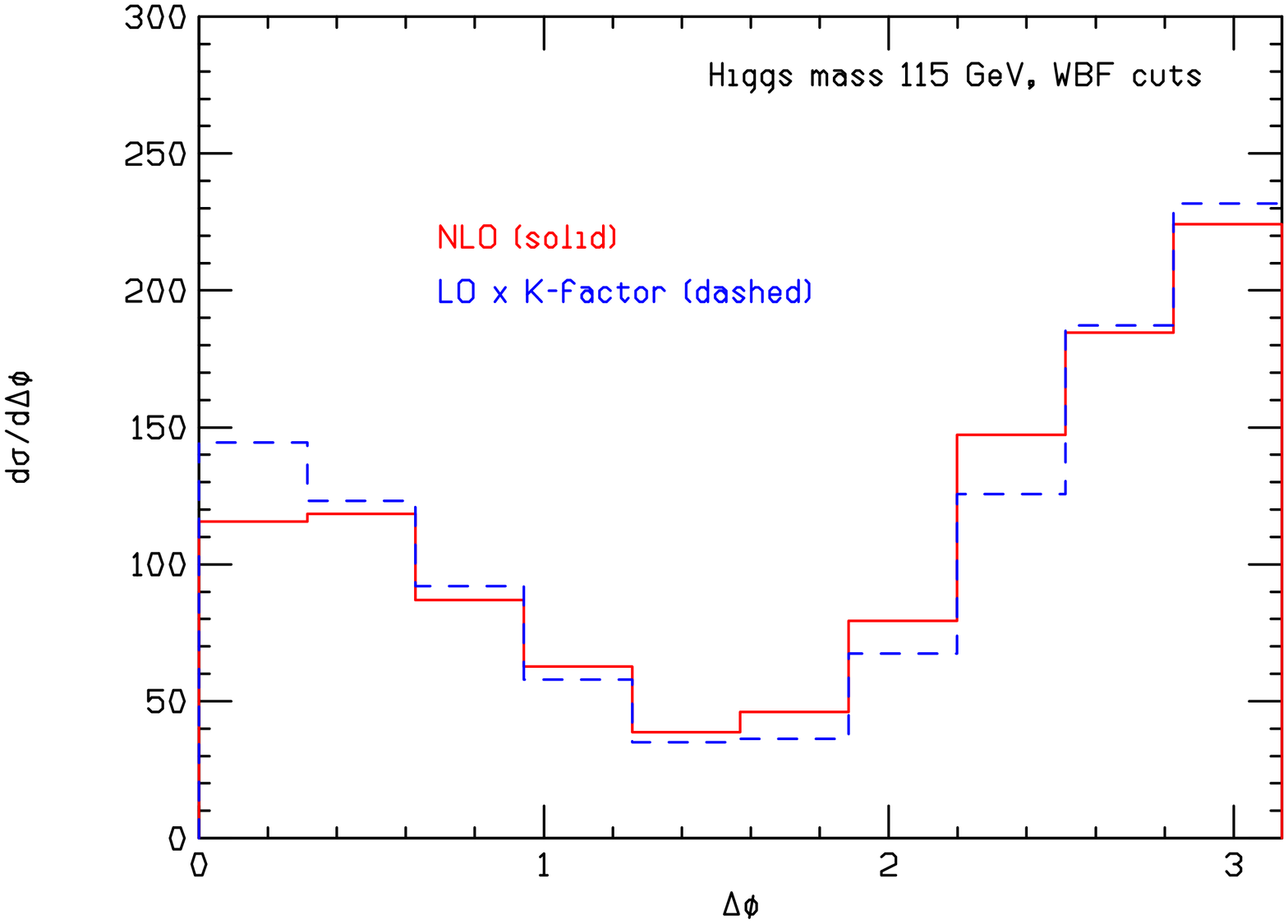,width=8cm,angle=90}
        \epsfig{file=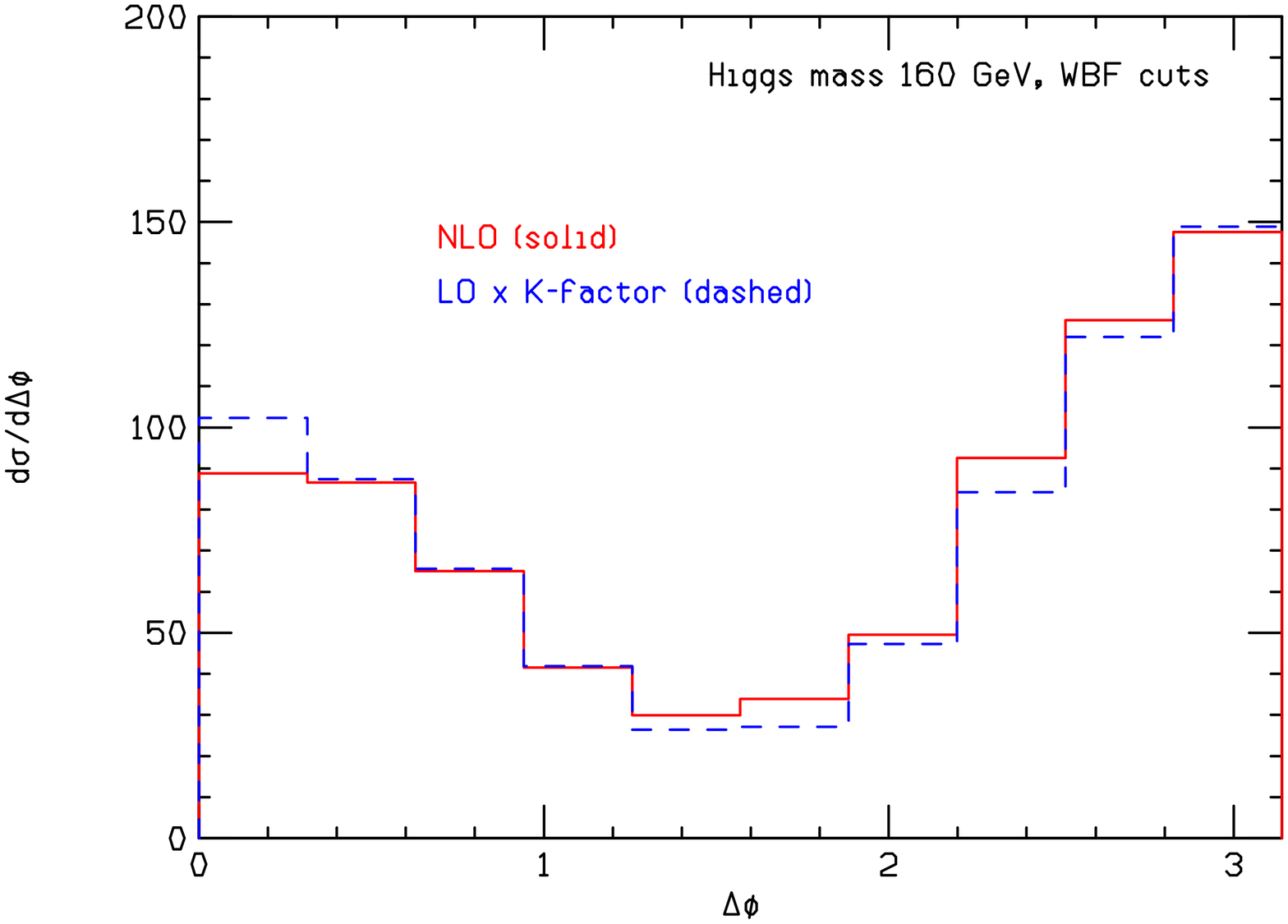,width=8cm,angle=90} 
        \caption{The azimuthal correlation of the two leading jets
          using the weak boson fusion cuts, for $m_H=115$~GeV (upper)
          and $m_H=160$~GeV (lower). The QCD process is calculated at
          NLO and LO, with the latter scaled up by the overall
          $K$-factor in this
        region.}%
        \label{fig:phi56wbf}}
Our results indicate that the shape of the lowest order distribution is
unchanged and therefore that the correlation survives the addition of
NLO QCD corrections.

\section{Conclusions}

We have presented a calculation of the production of a Higgs boson in
association with two jets at hadron colliders, performed in the limit
of large top mass and accurate to the next-to-leading order in the
strong coupling. Our results indicate that the effect of the QCD
corrections is modest and that the dependence of observables on the
factorization and renormalization scales is reduced. Furthermore we
find that the azimuthal correlation between the two leading jets, the
subject of recent parton shower based studies, is unchanged at NLO.

A major source of error in the extraction of the Higgs coupling to vector
bosons at the LHC is the theoretical uncertainty on the $H+2$~jet gluon
fusion process. Previous studies have estimated this uncertainty to be
$20\%$~\cite{Zeppenfeld:2000td} or $30\%$~\cite{Berger:2004pc}. Our
results suggest that the uncertainty due to the scale dependence alone may be
at least as large as this. Further study is necessary in order to extract
the coupling with greater confidence.

In the wider context, this calculation
represents the first full implementation of the semi-numerical
approach~\cite{Ellis:2005zh} for the virtual matrix elements.

\acknowledgments
We are grateful to Walter Giele for useful discussions. We would like
to thank Vittorio Del Duca, Alberto Frizzo and Fabio Maltoni for
providing us with their code for the evaluation of the $H+5$~parton
matrix elements and Daniel De Florian for comments on the manuscript.

\end{document}